\documentclass[journal=jctcce,manuscript=article,layout=traditional]{achemso}
\usepackage{helvet}
\usepackage{booktabs}

\usepackage[utf8]{inputenc}
\usepackage{url}
\usepackage{multirow}
\usepackage{wrapfig}
\usepackage{hyperref}
\usepackage{xcolor}

\definecolor{tmavaseda}{gray}{0.2}
\hypersetup{
pdfborder={0 0 0 [0 0]},
colorlinks=true,
urlcolor=teal,
linkcolor=tmavaseda,
citecolor=tmavaseda
}

\title{Sensitivity of peptide conformational dynamics in carbon nanotubes to directional mechanical forces}

\author{Felipe C. Nepomuceno}
\affiliation{Department of Physical Chemistry, University of Chemistry and Technology, Technická 5, 16628 Prague, Czech Republic}

\author{Michal H. Kolář}
\affiliation{Department of Physical Chemistry, University of Chemistry and Technology, Technická 5, 16628 Prague, Czech Republic}
\email{michal@mhko.science}

\begin{document}

\maketitle

\begin{abstract}
\singlespacing
In living organisms, proteins and peptides are often under the influence of mechanical forces, especially in confined spaces such as membrane channels, the ribosome exit tunnel, or the proteasome gate. Due to the directional nature of proteins as polymers with distinct ends, forces have the potential to influence protein conformational dynamics in a direction-dependent manner. In this study, we employed force-probe molecular dynamics simulations to investigate the impact of pulling a peptide through a confined environment versus pushing it in the same direction. Our model involves a carbon nanotube and one of three decapeptides with varying amino acid sequences. The simulations reveal that the difference between pulling the C-terminus and pushing the N-terminus is relatively minor when considering the conformational ensembles of the peptides. The loading rate of the force probe and the amino acid sequence of the peptide play a more important role. However, the application of force to the peptide significantly influences the relative motion of the peptides with respect to the nanotube. Pulling the peptide results in a smoother translocation through the nanotube compared to pushing, although the internal conformational dynamics of the peptide add complexity in either case. Our findings shed light on how short peptides navigate confined spaces within the cellular environment, emphasizing the importance of force-probe simulation studies in understanding these processes. 
\end{abstract}

\singlespacing

\section{Introduction} 

Proteins are ubiquitous biopolymers that are involved in almost all processes in cells. During their life, proteins exist in a crowded environment of other biomolecules in the cytosol. Temporarily, proteins also occur spatially confined in cavities, pores, or channels. For example, during the translation elongation phase in the ribosome, the nascent polypeptide chain passes through the approximately 10\,nm long ribosome exit tunnel which has a diameter varying between 0.8 and 1.5\,nm \cite{Voss06,DaoDuc19a}. A protein intended for export to the extracellular space must be transported through the translocon channel in the cell membrane, the pore of which is narrower than 0.5\,nm \cite{Berg04}. Before degradation by proteasomes, proteins must be unfolded and inserted into the proteasome cavity through a gate with a diameter less than 1 nm \cite{Rabl08}.

It is well established that the confined space alters the conformational dynamics of proteins. Restricting the conformational heterogeneity of the unfolded state accelerates protein folding and stabilizes the folded conformation \cite{Ping03, Zhou08, Lucent07, Mittal08}. Studies on shorter proteins or peptides that are more relevant to the ribosome tunnel or the translocon channel revealed that the effect of confinement is difficult to generalize. Simulations of a coarse-grained off-lattice model suggested an entropic stabilization of $\alpha$-helices in a cylinder\cite{Ziv05}. On the other hand, all-atom molecular dynamics (MD) simulations proposed a destabilizing effect of carbon nanotube on $\alpha$-helices compared to free peptides in solution \cite{Sorin06}. O'Brien \emph{et al.} showed that the stability of $\alpha$-helices in cylinders is largely affected by the diameter of the cylinder and the amino acid sequence of the polypeptide \cite{OBrien08}. This conclusion was later confirmed by more advanced MD simulations \cite{Suvlu17}.

One of the many questions related to the processes in confined spaces or constrictions concerns the nature of the forces that act on the peptides out of equilibrium. Is the peptide exported from the cell pulled or pushed through the translocon pore? Is there a pushing force acting on the nascent polypeptide or is it just pulling that drives the peptide through the ribosome tunnel towards the exit? How does an unfolded protein overcome the proteasome gate?

MD simulations have proven useful in studies of force-induced processes involving biomo\-le\-cules. In computer simulations known as targeted MD \cite{Schlitter94}, steered MD \cite{Isralewitz01}, or force-probe MD \cite{Grater05,Grubmuller05} (fpMD), an external force is applied to drive the system out of equilibrium in a finely controlled manner. These techniques have also been used to characterize peptide behavior in cylindrical spaces such as the ribosome exit tunnel.

In an early attempt, Ishida and Hayward studied the principles of spontaneous translocation of nascent polyalanine through the exit tunnel \cite{Ishida08}. Because the translocation takes longer than their simulation times of 2 ns, they used fpMD to speed it up. They pulled the N-terminus of the nascent polypeptide toward the exit of the tunnel and identified two pathways through which the nascent polypeptide overcomes the narrowest part of the exit tunnel.

Another fpMD study focused on the nascent polypeptide SecM, which causes translational arrest \cite{Nakatogawa01} by stabilizing an unproductive conformation of two tRNAs within the ribosome \cite{Gersteuer24}. FpMD simulations with a pulling force acting on the N-terminus of the nascent polypeptide were used to mimic the action of the SecA translocon motor, which can relieve the arrested state of the ribosome by displacing SecM from the arresting conformation \cite{Butkus03}. Simulations helped characterize a series of molecular events in the exit tunnel that followed force-induced SecM conformational changes \cite{Gersteuer24}. More recently, Zimmer \emph{et al.} used fpMD to release ribosomes stalled also by SecM and VemP arresting peptides \cite{Zimmer21}. By pulling the N-termini of the arresting peptides toward the tunnel exit, they identified multiple energy barriers along the translocation pathway, although the initial SecM conformation was inaccurate, according to more recent findings \cite{Gersteuer24}.

In the experimental approach, the pulling forces are applied to the nascent polypeptides by means of AFM or optical tweezers. For example, it was confirmed that the pulling force required to release the SecM-stalled ribosome can be generated \emph{in vivo} by cotranslational folding of the nascent polypeptide \cite{Goldman15}. Short peptides were also studied under external pulling force when attached on hydrophobic and hydrophilic surfaces \cite{Erbas12}. Using atomistic fpMD simulations, it was established that the friction force is proportional to the number of hydrogen bonds between the peptide and the hydrophilic surface and the pulling velocity.

To the best of our knowledge, no pushing motion has been employed to study peptide translocations. In fact, it is unclear whether pushing a peptide affects the conformational dynamics of spatially confined peptides differently from pulling them. There are hints that the direction of the force is important in complex biomolecular systems. For example, mechanical forces of pulling and pushing were applied using AFM to the Piezo1 membrane receptor of living animal cells in the presence or absence of extracellular matrix proteins \cite{Gaub17}. The receptor sensitivity was shown to depend both on the matrix proteins and on the direction of the force. Furthermore, the mechanism of export of peptides through the translocon can involve steps with both pushing and pulling forces acting on the peptide \cite{Tomkiewicz07}. The nature of the forces that act on the nascent polypeptides before they reach the tunnel exit remains unclear. In later stages of protein elongation, a pull generated by protein folding certainly facilitates the nascent polypeptide translocation \cite{Cassaignau20}.

Here, we study how the direction of the mechanical force affects the conformation dynamics of spatially confined model peptides. To simplify the problem, we model the confined space using a carbon nanotube (CNT), with a diameter roughly the same as that of the ribosome exit tunnel. The tunnel creates an environment with heterogeneous dielectric behavior \cite{Lucent10b}. Furthermore, the geometry of the ribosome exit tunnel is rugged \cite{Voss06}. Nonetheless, using the CNT as the first approximation, our approach allows us to focus our attention on the peptide. In this way, we removed any directionality that could arise when an asymmetric or rough confined environment is used. Through extensive fpMD simulations, we address the question of what is the difference between pulling and pushing the peptide through the CNT.

\section{Methods}
\label{sec:methods}

\subsection{Simulated systems}

\begin{figure}
    \centering
    \includegraphics{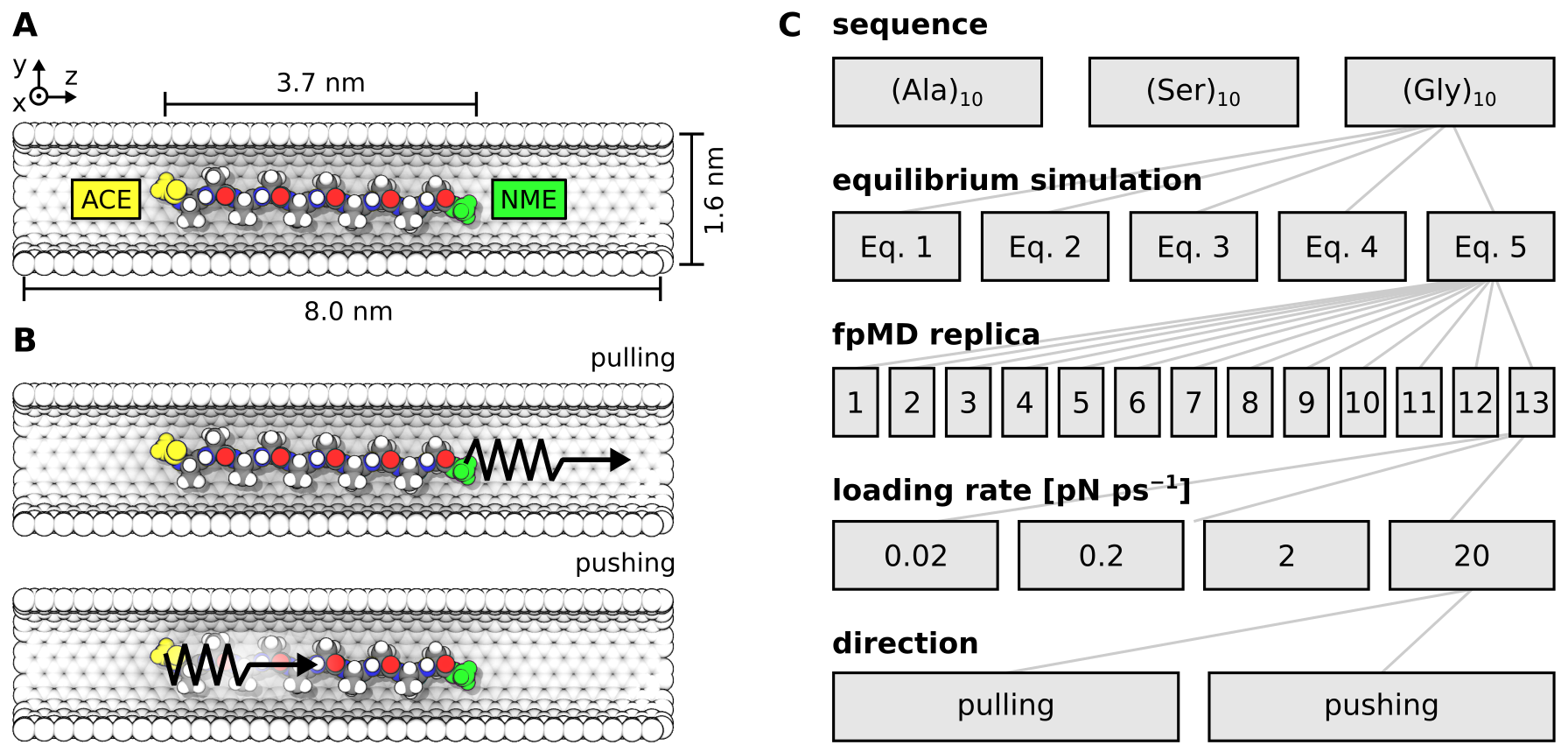}
    \caption{A) An overview of the poly-alanine in the carbon nanotube. The terminal acetyl (ACE) and N-methyl (NME) residues are labeled in yellow and green, respectively. The peptide is color-coded by atoms -- C: gray, O: red, N: blue, H: white. Water molecules were omitted for clarity. B) The force probe position for the pulling and pushing non-equilibrium simulations. C) Schematic illustration of the simulations performed. Note, that for the loading rate of 0.02\,pN\,ps$^{-1}$, only 5$\times$7 replicates were simulated, instead of 5$\times$13 as for the loading rates 0.2, 2, and 20\,pN\,ps$^{-1}$.} 
    \label{fig:mdOverview}
\end{figure}

We examined three decapeptides: poly-alanine, poly-serine, and poly-glycine. The peptides were placed in their fully extended conformations in an armchair CNT with a diameter of 1.6\,nm, roughly corresponding to the narrower parts of the ribosome exit tunnel. The length of the CNT was set to 8\,nm to allow the peptide to adopt an extended conformation without any contact with its periodic image. The peptide was oriented with its N-terminus in lower Z coordinates and its C-terminus in higher Z coordinates. An acetyl group (ACE) was attached to the N-terminus and a methyl amide group (NME) to the C-terminus of each peptide (Fig.\,\ref{fig:mdOverview}). For the preparation of the systems, we used VMD\cite{Humphrey96} and PyMOL\cite{DeLano02}.

The systems were placed in an orthogonal triclinic simulation box such that the CNT axis coincided with the Z axis of the Cartesian space. Periodic boundary conditions were used in all three directions. The box dimensions in the X and Y directions were 3.0\,nm. In the Z direction, the box dimension was chosen so that the CNT was infinite. In this way, we discarded any effects related to the finite size of the CNT or the interface between the CNT end and water. The box and CNT were filled with water.

The peptides were described by the CHARMM36m force field, which is optimized for folded and disordered proteins and peptides \cite{Huang17}. The CNT was modeled as a set of CHARMM36m particles of type C. For simplicity, no bonded, angular, or torsion parameters were used within the CNT. Instead, to prevent the collapse of the structure, harmonic position restraints with a force constant of 50\,000\,kJ\,mol$^{-1}$nm$^{-2}$ were used. Lennard-Jones interactions were excluded for all pairs of CNT carbon atoms closer to 5\,nm. For water, the rigid TIP3P model was used \cite{Berendsen87, Huang17}.

Coulomb interactions were described using the particle-mesh Ewald method \cite{Darden93} with a grid spacing of 0.12\,nm and a cutoff of 1.2\,nm. A potential-shift-verlet modifier was used. Lennard-Jones interactions were truncated at 1.2\,nm using a force-switch modifier with a switch distance of 1.0\,nm. Covalent bonds involving hydrogens were converted to constraints and treated by the LINCS \cite{Hess08} algorithm of the order of 6.

\subsection{Equilibrium simulations}
The potential energy of the systems was minimized using the steepest descent algorithm until reaching computer precision (roughly in 1000 steps). The solvent was then heated to 310\,K in a 2-ns long MD simulation using v-rescale temperature coupling \cite{Bussi07}. The initial atomic velocities were randomly drawn from the Maxwell--Boltzmann distribution at 5\,K. The pressure was equilibrated in a 2\,ns simulation using the semi-isotropic Berendsen barostat \cite{Berendsen84} with a reference pressure of 1\,bar and system compressibility of 4.5$\cdot10^{-5}$\,bar$^{-1}$.

To sample equilibrium conformations of the peptides in the CNT at 310\,K, for each peptide, we performed five independent NVT simulations that differed in the initial velocities. Each simulation was 1\,$\mu$s long and used a time step of 2\,fs. System configurations were saved every 1\,ps. Simulations were run using the GROMACS 2019 package \cite{Abraham15}.

\subsection{Force-probe pulling and pushing}

To distinguish between pulling and pushing motions throughout the CNT, we performed two sets of simulations for each peptide. The molecular setup is depicted in Fig.\,\ref{fig:mdOverview}B. We \emph{pulled} the peptide with a force probe attached to the C-terminus and moved the probe in the positive sense of the Z direction. On the other hand, we \emph{pushed} the peptide using a force probe attached to the N-terminus and also moving it in the positive sense of the Z direction.

The force was applied in the Z direction only, so the termini were free to move in the XY plane. The force constant of the harmonic probe (defined as pull-coord-type=umbrella in GROMACS) was set at 602.4\,kJ\,mol$^{-1}$. To evaluate the role of the loading rate (the product of the velocity of the probe and the force constant), the force probe was moved with four different constant velocities: 20, 2, 0.2, and 0.02 nm\,ns$^{-1}$. As the reference group for the force probe, the center of mass of the CNT was used. The velocity of the probe and the force constant resulted in four different loading rates: 20, 2, 0.2, and 0.02 pN\,ps$^{-1}$. All fpMD simulations were run until the total displacement of the peptides reached 10\,nm in the direction in which it was moving. Therefore, the simulations had different lengths depending on the loading rate, that is, 0.5\,ns, 5\,ns, 50\,ns, and 500\,ns. A summary of the simulations performed is depicted in Fig.\,\ref{fig:mdOverview}C.

\subsection{Obtaining initial configurations for fpMD}
\label{sec:filtering}

We carried out 65 independent fpMD simulations (replicates) for each system and type of motion with loading rates of 20, 2, and 0.2\,pN\,ps$^{-1}$. For the loading rate of 0.02\,pN\,ps$^{-1}$ we performed 35 independent simulations.

\begin{figure}
    \centering
    \includegraphics{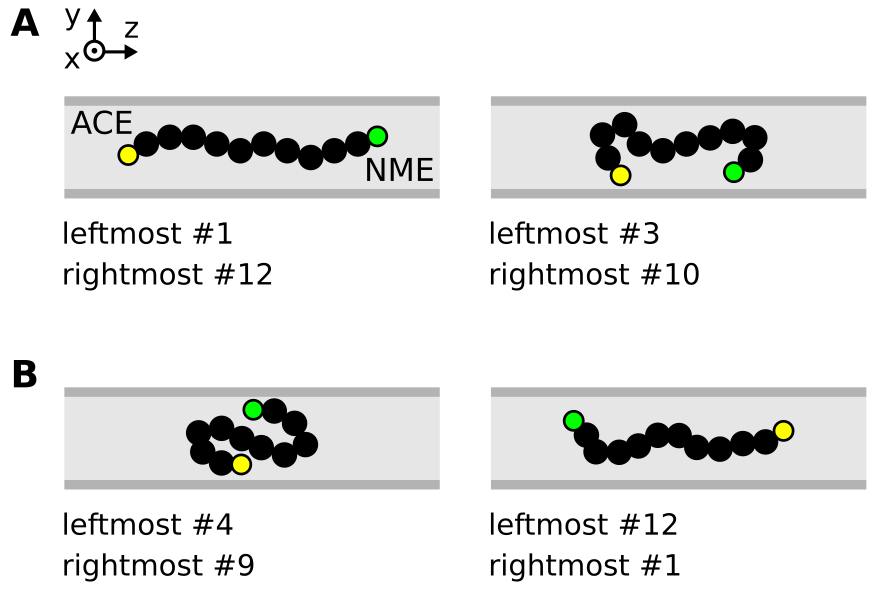}
    \caption{A schematic representation of the peptide conformations highlighting various termini positions. Amino acid residues are black, N-terminal acetyl group (ACE) in yellow, C-terminal N-methyl group (NME) in green.  Residue number of the leftmost and rightmost residues are shown. A) Examples of acceptable conformations for the fpMD. B) Examples of rejected conformations.}
    \label{fig:bends-scheme}
\end{figure}

The initial configurations for each replicate were obtained from the equilibrium simulations. However, during equilibrium simulations, the peptides temporarily folded or bent to form hairpin-like structures inside the CNT, so it was unclear whether the fpMD would represent a pulling or pushing motion (Fig.\,\ref{fig:bends-scheme}). Moving a hairpin-shaped peptide through the CNT would mean pushing one part of the peptide and pulling the other. To avoid these ambiguous situations, we considered only those equilibrium trajectory frames in which the peptide was in a sufficiently extended conformation. A frame was suitable for fpMD when the rightmost residue of the peptide (\emph{i.e.} with the highest Z coordinate) was one of the three C-terminal residues and the leftmost residue of the peptide (\emph{i.e.} with the lowest Z coordinate) was one of the three N-terminal residues. This filtering setup allowed the peptide termini to be slightly bent, avoiding extensive elimination of the equilibrium trajectory frames.

In each 1\,$\mu$s trajectory, only the last 900\,ns was used to generate the initial configurations for the fpMD simulations. During the first 100\,ns, the peptides were left to equilibrate.

\subsection{Analysis}

We analyzed the equilibrium and fpMD simulations separately. If not stated otherwise, the last 900\,ns of the equilibrium trajectories were used. For the fpMD simulations, we used the last 50\% of the simulation time, while keeping the number of frames the same. Thus, the data for the slower fpMDs represent longer time intervals than for the faster fpMDs. For various properties of the simulated system, we calculated their change upon application of mechanical force. The change in property $X$ was defined as $\Delta X = \langle X\rangle^{\mathrm{fpMD}}_{50\%} - \langle X\rangle^{\mathrm{eq}}_{900\,\mathrm{ns}}$.

We calculate the end-to-end distance (E2E) as the Euclidean distance between the carbon atom of the ACE group and the nitrogen atom of the NME group. The number of hydrogen bonds (H-bonds) and hydrogen bond-like pairs within the peptide was obtained using the \emph{gmx hbond} tool of the GROMACS package. The H-bond was defined as an interatomic contact between hydrogen and an acceptor atom (N or O) shorter than 0.35\,nm with the angle X--H--acceptor between 150$^\circ$ and 180$^\circ$. The pair was defined in the same way, but without any angular dependence. The solvent accessible surface area (SASA) was calculated using the \emph{gmx sasa} tool with the default probe radius of 0.14\,nm.

The position of the peptide within the CNT in the $xy$ plane was characterized using the cylindrical radius  $r$, \emph{i.e.} the radial distance of the C$_\alpha$ atoms from the CNT axis. The probability density function (pdf) of the C$_\alpha$ positions was calculated as a function of the cylindrical radius $r$.

The equilibrium and fpMD simulations were analyzed with custom scripts using tools from the GROMACS package \cite{Abraham15} and the MDAnalysis Python library \cite{Michaud-Agrawal11, Gowers16}.

\section{Results and discussion}

\subsection{Equilibrium simulations and their filtering}

\begin{figure}
    \centering
    \includegraphics{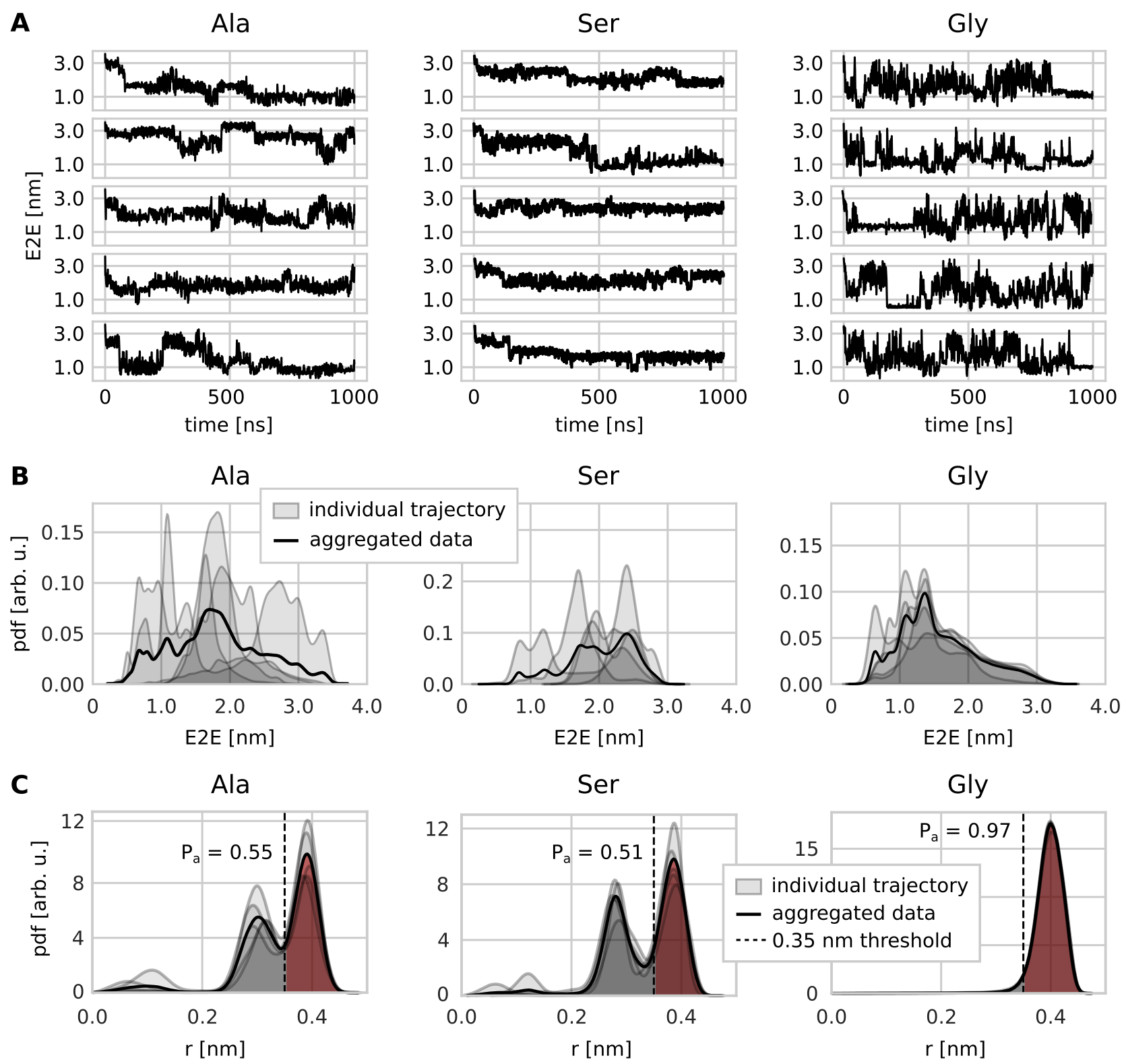}
    \caption{Analysis of the equilibrium trajectories. A) End-to-end distance (E2E) as a function of simulation time shown for five independent trajectories of each simulated polypeptide sequence. B) Probability density function (pdfs) of E2E calculated from individual trajectories (gray areas) and from aggregated data (black line). C) Pdfs of C$_\alpha$ positions as a function of cylindrical radius $r$. The red integrated area of pdf with $r>$0.35\,nm represents the polypeptide propensity $P_a$ to associate with the CNT wall.}
    \label{fig:equil}
\end{figure}

Using unbiased MD simulations, we generated equilibrium ensembles of three decapeptides in the CNT. Fig.\,\ref{fig:equil}A shows the peptide E2E as a function of the simulation time and Fig.\,\ref{fig:equil}B shows the respective pdfs. All of the pdfs overlap regardless of the peptide sequence. They span a wide range of less than 1\,nm to more than 3\,nm, which means that the three peptides studied are conformationally diverse. In the CNT, they may adopt a compact conformation with low E2E or be quite extended.

The high conformational diversity of short peptides is also typical in bulk solution, as shows, for example, for Ala$_{10}$ \cite{Hazel14}. For globular proteins, the situation is more complicated: on the one hand, confinement destabilizes the unfolded state of the protein; on the other hand, the solvent structure differs under confinement versus bulk, which destabilizes the folded state \cite{Lucent07}. The effect of solvent on the folding equilibrium was also observed for a helical peptide \cite{Sorin06, Suvlu21}.

We observed certain differences among the sequences in CNTs. The narrowest pdf, representing a more homogeneous conformational ensemble, was observed for poly-Ser. Serine is an amino acid with a small and polar side chain capable of H-bonding, which promotes protein disorder \cite{Romero01}. At the beginning of the simulations, the fully extended conformations of all peptides had an E2E of 3.7\,nm. Poly-Ser typically adopted conformations with E2E lower than 3.0\,nm. Slightly wider pdfs were observed for poly-Ala and poly-Gly, where E2E almost reached the maximum value. These extremely extended conformations are probably prohibited by the poly-Ser intramolecular H-bonds. On average, we observed about 3 H-bonds, for poly-Ser, and less than 1 H-bond for poly-Ala and poly-Gly. Poly-Gly, which has the smallest side chains -- a single hydrogen atom -- showed fast and extensive fluctuations of E2E, compared to poly-Ala and poly-Ser, where the fluctuations of E2E were much smaller.

The positions of the peptides also differ in the $xy$ plane within the CNT. Fig.\,\ref{fig:equil}C shows the pdf of the C$_\alpha$ positions as a function of the cylindrical radius $r$. We integrate pdfs for $r>$ 0.35\,nm and obtain propensities $P_a$ of the peptides to associate with the CNT wall. The propensities are 0.55, 0.51, and 0.97 for poly-Ala, poly-Ser, and poly-Gly, respectively. The pdf of poly-Gly has only one maximum near the CNT wall. This means that poly-Gly is attached to the wall throughout the simulations and most of its C$_\alpha$ atoms are within 0.35\,nm of the CNT wall. Poly-Ala and poly-Ser show a multimodal distribution with a non-negligible portion of the time spent dissociated from the CNT wall closer to the CNT axis. In fact, poly-Ser has the lowest $P_a$. Clearly, the hydroxyl groups of poly-Ser allow for a better solvation of the peptide, thus the lower $P_a$. The overlapping pdfs justify that the positions of the peptide are well-converged across the independent replicas.

\begin{figure}[bt]
    \centering
    \includegraphics{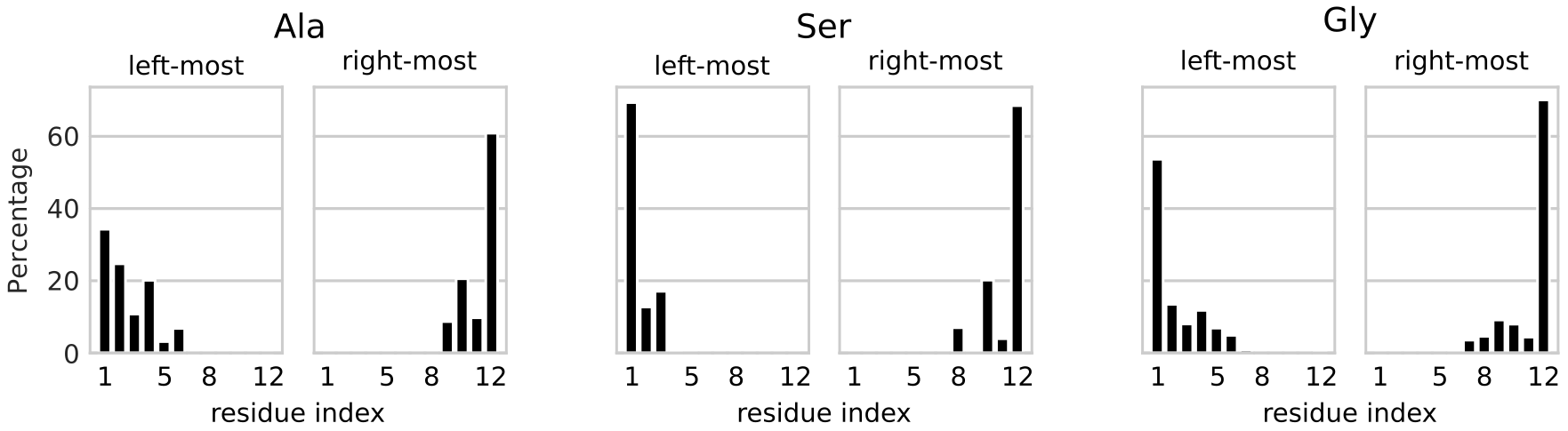}
    \caption{Residues identified as leftmost or rightmost extremes in the Z-direction of the simulation box. The bars stand for a percentage of frames, where the given residue occurred as extreme. Note that the first and last residues are the acetyl and N-methyl caps, respectively.}
    \label{fig:elimFrames}
\end{figure}

For fpMD, we filtered the equilibrium trajectory frames in which the peptides appeared to be bent. The percentage of frames eliminated was 36\%, 8\%, and 42\%, for poly-Ala, poly-Ser, and poly-Gly, respectively. Fig.\,\ref{fig:elimFrames} shows the outermost residues, identified as extremes in the Z direction, in the equilibrium simulations. In this respect, poly-Ser remained oriented well within the CNT with only slight bends of its termini. In contrast, because of its high flexibility, poly-Gly bent extensively, so a large number of frames could not have been used for fpMD.

\subsection{Impact of the force on the peptide conformations}

To address the impact of the force on the peptide conformational ensembles, we compared the second halves of the fpMD trajectories with the equilibrium simulations. First, we calculated $\Delta$E2E as the difference between the E2E averaged over the last 50\% of the fpMD simulations and the E2E averaged over the last 90\% of equilibrium simulations. Therefore, the negative values of $\Delta$E2E mean a shrinkage of the peptide during fpMD, and the positive values represent the extension of the peptides.

\begin{figure}[bt]
    \centering
    \includegraphics{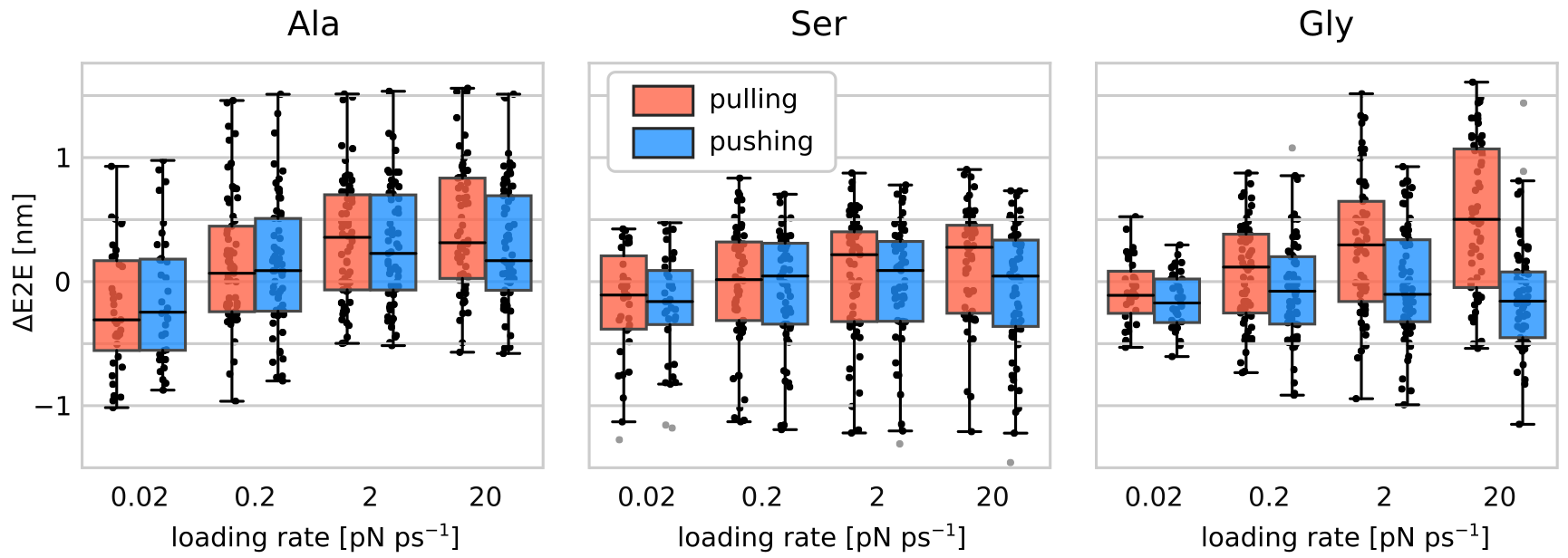}
    \caption{$\Delta$E2E as the difference between the E2E averaged over the last 50\% of fpMD simulations and the E2E averaged over the last 90\% of equilibrium simulations. Each dot stands for an independent fpMD trajectory. The boxes represent interquartile ranges with the mean values indicated. The whiskers stand for the minimum/maximum values excluding the outliers (gray circles).}
    \label{fig:E2Ediff}
\end{figure}

The $\Delta$E2E is plotted in Fig.\,\ref{fig:E2Ediff} for four loading rates of pulling and pushing fpMD. $\Delta$E2E from individual fpMD trajectories span a wide range of values; within the simulation time of a single fpMD simulation, the peptide can extend by up to 2\,nm, or shrink by 1\,nm regardless of the direction of mechanical force.

On average, the observed trends depend on the peptide sequence, and less on the loading rate. Poly-Gly is the most susceptible to mechanical forces, likely due to its high flexibility. The difference in mean $\Delta$E2E between pulling and pushing increases with loading rate. For the slowest pulling, the $\Delta$E2E is around 0\,nm, which indicates that the peptide conformation is not affected by mechanical force. For the highest loading rate of pulling studied, poly-Gly is elongated by mechanical force by approximately 0.5\,nm on average. The spread of $\Delta$E2E values is so large that even for the highest loading rate, poly-Gly can become shorter (negative $\Delta$E2E) during the pulling fpMD. The effect of mechanical force during pushing is less apparent. Poly-Gly shrinks slightly on average for high loading rates.

The trend in poly-Ala $\Delta$E2E is similar to that of poly-Gly, but weaker. With an increasing loading rate of the pull, the peptide extends slightly on average. The spread of $\Delta$E2E of poly-Ala is large regardless of the loading rate and type of applied force. Even for pushing simulations, poly-Ala may elongate and apart from the lowest loading rate the mean values of $\Delta$E2E are slightly positive. This means that the intrinsic dynamics of poly-Ala in CNT is not intensely affected by mechanical forces.

The smallest effect of pulling and pushing on $\Delta$E2E is observed for poly-Ser. On average, $\Delta$E2E remains zero for all loading rates and types of motion. However, $\Delta$E2E spans a wide range of values, similar to the other peptide sequences. This reflects the intrinsic properties of the peptide in the CNT observed also in the equilibrium simulations, rather than the effect of an external mechanical force. 

To better characterize the conformational ensembles under mechanical forces, we calculated $\Delta$SASA, which describes the changes in peptide solvation. Furthermore, because the peptide dynamics depends on the intramolecular interactions, we calculated number of pairs. Their change related to the externa force, $\Delta$pairs was calculated in the same way as $\Delta$E2E. We calculated

\begin{figure}[tb]
    \centering
    \includegraphics{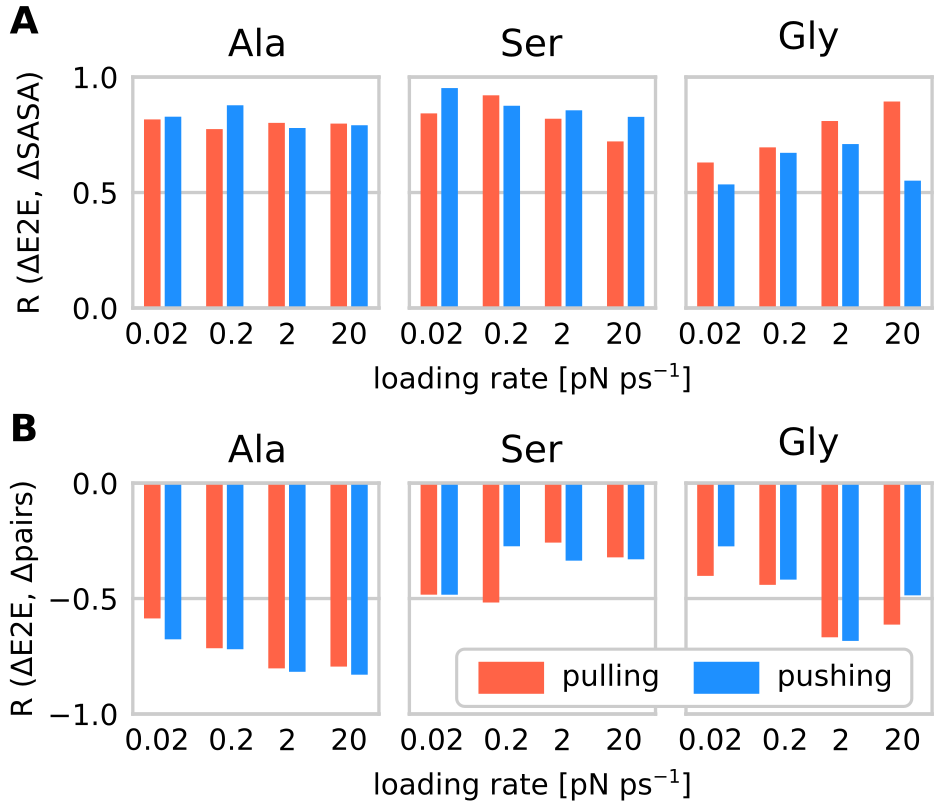}
    \caption{A) Pearson correlation coefficient $R$ between the change of the end-to-end distance ($\Delta$E2E) and the change in the solvent-accessible surface area ($\Delta$SASA). B) Pearson correlation coefficient $R$ between $\Delta$E2E and the change in the number of intramolecular contact pairs ($\Delta$pairs).}
    \label{fig:correlations}
\end{figure}

We performed a correlation analysis for $\Delta$E2E and $\Delta$SASA and for $\Delta$E2E and $\Delta$pairs. The correlation coefficients $R$ obtained are shown in Fig.\,\ref{fig:correlations}.

We observed a strong correlation between $\Delta$E2E and $\Delta$SASA for poly-Ala and poly-Ser with a typical $R$ of approximately 0.8. This was independent of both the type of applied force and the loading rate. This means that the elongation of the peptide as a result of the mechanical force is correlated with the increase in the SASA. Poly-Ala and poly-Ser become more exposed to solvent when they are elongated.

A different behavior was observed for poly-Gly. First, $R$ between $\Delta$E2E and $\Delta$SASA was lower than for the other two sequences, and second, it depended more strongly on the type of applied force and the loading rate. For pushing, $R$ was smaller than for pulling. Moreover, with the loading rate of pushing, $R$ became lower.

The internal dynamics of the polypeptides depends on intramolecular interactions. In our simulations, the average number of H-bonds within the peptides was approximately 0.4, 3.2, and 0.2 for poly-Ala, poly-Ser, and poly-Gly, respectively. Because of the low number of H-bonds, their changes during pulling or pushing showed a poor signal-to-noise ratio. Consequently, we concentrated on the pairs, as their average number count was approximately tenfold greater than that of H-bonds. We found that the force-induced change in the number of pairs $\Delta$pairs was anticorrelated with $\Delta$E2E. The negative $R$ goes in line with the notion that when the peptide is elongated by external force, the number of intramolecular contact pairs decreases. This behavior is common for all three sequences and does not depend on the type of applied force. While in poly-Ser and poly-Gly, $R$ is also independent of the loading rate, we see some dependence on the loading rate in the case of poly-Ala. For higher loading rates, $R$ becomes more negative, reaching about --0.8 for the loading rate of 20 pN\,ps$^{-1}$.

\subsection{Impact of the force on the peptide position}

We analyze how the external force affects the position of the peptide within the CNT. The first and obvious effect is the translational motion of the peptide with respect to the CNT in the direction of the applied force. The simulation length and the velocity of the force probe ensured that the peptides traveled 10 nm in the Z-direction. This motion was consistent for all the sequences and loading rates studied.

Although the force probe was attached to the peptide termini and moved in the Z direction, the intrinsic dynamics of the peptide caused it to bend, as also observed in the equilibrium MD simulations (Fig.\,\ref{fig:bends-scheme}). Consequently, for pulling, the rightmost residue was not always the terminal N-methyl but some \emph{inner} amino acid residue. Similarly, for pushing, the leftmost residue was not always the terminal acetyl. 

\begin{figure}[tb]
    \centering
    \includegraphics{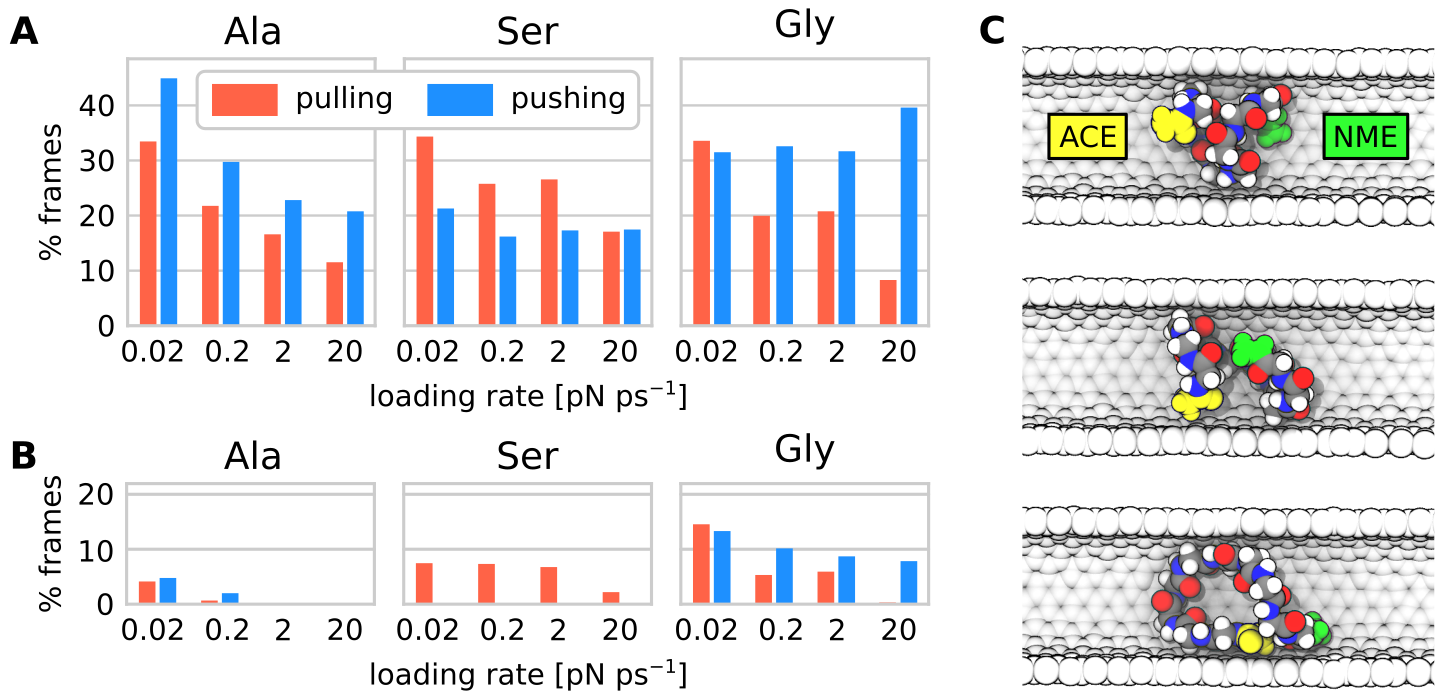}
    \caption{Peptide bends during fpMD. A) Percentage of frames in which the leftmost or rightmost residue was not the acetyl (ACE) or N-methyl group (NME). B) Percentage of frames where at least three residues are bent. C) Representative snapshots of simulations pushing poly-Gly, where the termini are well oriented without any bend (top panel), three residues are bent (middle panel) and the peptide is almost flipped (bottom panel).}
    \label{fig:bends}
\end{figure}

To characterize how often the peptide bent when pushed or pulled, we calculated the percentage of fpMD frames in which the leftmost or rightmost residues were not acetyl or N-methyl groups, respectively. Fig.\,\ref{fig:bends} shows that this situation is quite frequent and depends both on the type of applied force and the loading rate. For pulling, the bends appeared less often than for pushing, regardless of the peptide sequence. In addition, the higher the loading rate, the less frequently the peptide bends.

Next, we calculated the percentage of fpMD frames in which the peptide was bent more dramatically. Namely, we considered frames in which the leftmost or rightmost residue was not any of the first three or last three residues. We used this approach to filter the equilibrium simulations; see Section \ref{sec:filtering}. Compared to the previous criterion with a single bent residue, the percentage of frames with at least three bent residues was much lower. For poly-Ser, we did not observe any occurrence of such a situation, poly-Ala appeared bent in less than 5\% frames and mostly for the lowest loading rate. Poly-Gly, on the other hand, was bent especially during pushing fpMD simulations. Fig.\,\ref{fig:bends}B shows three representative snapshots of simulations pushing poly-Gly, where the termini are well oriented without any bend (top panel), at least three residues are bent (middle panel), and the peptide is almost flipped (bottom panel).

\begin{figure}[tb]
    \centering
    \includegraphics[width=\textwidth]{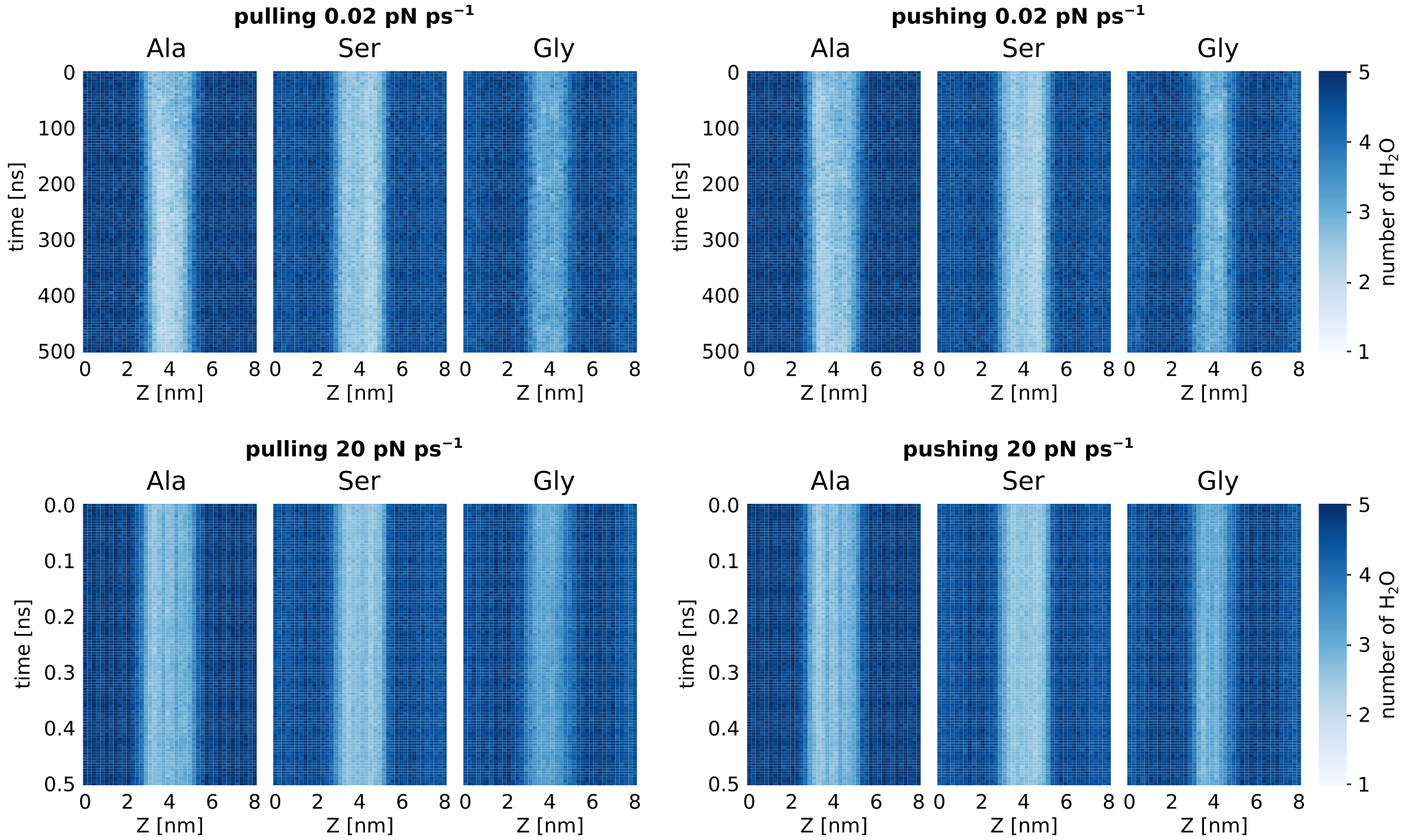}
    \caption{Solvation patterns represented by a time series of a number of molecules in a particular segment of the CNT. Color coded from white (low nomber of water molecules) to blue (high number of water molecules). Data for the lowest and highest loading rates are shown.}
    \label{fig:solvation}
\end{figure}

As the peptides traverse the CNT, the solvation pattern around the peptide may undergo remodeling. For instance, water molecules can infiltrate the compact peptide structure, or interactions between the CNT and the peptide may be replaced by interactions between water and the peptide. To analyze this phenomenon, we calculated the number of water molecules along the CNT as a function of simulation time (Fig.\,\ref{fig:solvation}). The simulation boxes were aligned relative to the peptide, and the water profiles were averaged across the fpMD trajectories.

The solvation of the peptide is sequence-dependent. In agreement with the side-chain size, poly-Gly exhibits the highest number of surrounding water molecules within a given segment of CNT, while poly-Ala is associated with significantly fewer water molecules. No obvious changes in solvation were observed during the fpMD simulations of poly-Gly and poly-Ser. However, poly-Ala showed a distinct change in its solvation pattern during pulling and pushing, as evidenced by the narrowing of the bright region over time (Fig.\,\ref{fig:solvation}). The effect of external force was observed only for the slowest simulations. In the faster simulations, \emph{i.e.} with a higher loading rate, we observed a steady solvation pattern. On average, the water molecules were not rearranged enough fast.

\begin{figure}[tb]
    \centering
    \includegraphics{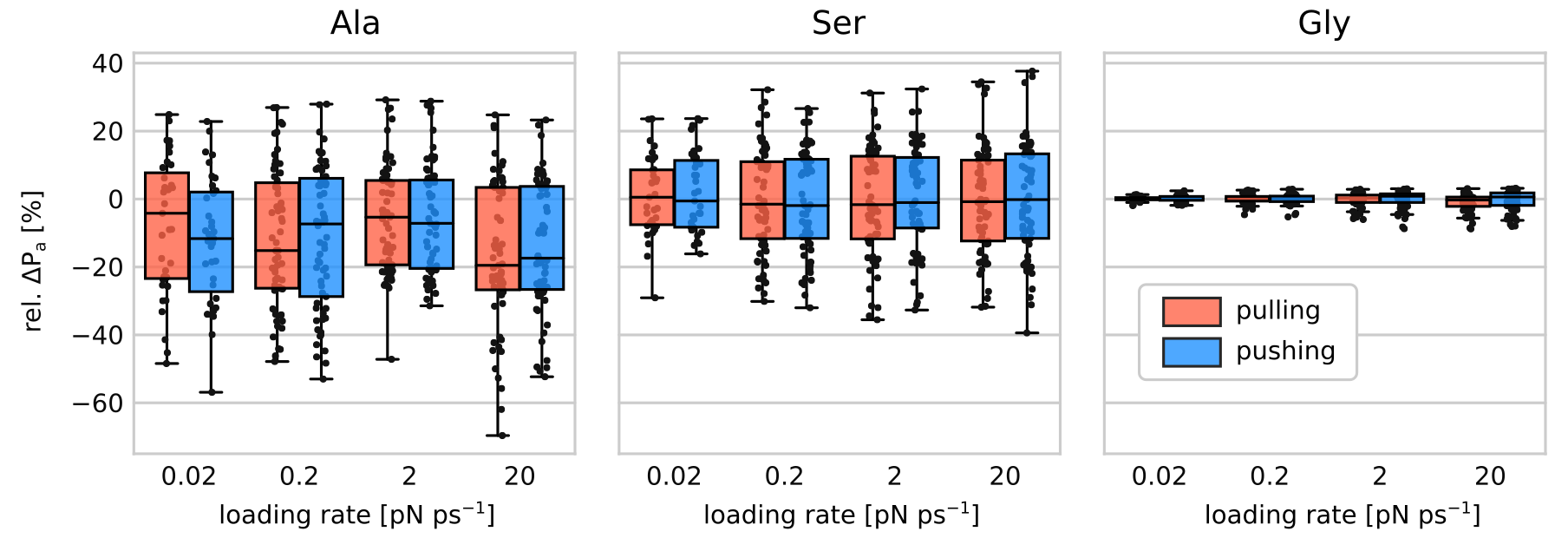}
    \caption{A relative change of the peptide propensity $P_a$ to associate with the CNT wall caused by the mechanical force. Each dot stands for an independent fpMD trajectory. The boxes represent interquartile ranges with the mean values indicated. The whiskers stand for the minimum/maximum values excluding the outliers.}
    \label{fig:association}
\end{figure}

Changes in peptide solvation are reflected in the propensities of peptides to associate with tunnel walls, $\Delta P_a$. By normalizing with the value of $P_a$ from the equilibrium simulations, we obtained a relative change of $P_a$ caused by the mechanical force (Fig.\,\ref{fig:association}). Consequently, the negative values of rel.$\Delta P_a$ represent the situations in which the mechanical force caused the dissociation of the peptide from the CNT wall.
 
On average, the mechanical force applied to the peptides causes dissociation from the CNT walls; the mean values of rel.$\Delta P_a$ are negative. The range of values of individual fpMD simulations depends strongly on the sequence. For poly-Gly, rel.$\Delta P_a$ values are close to zero, except for the highest loading rate, which means that the propensity of poly-Gly to associate with the CNT wall remains the same even after the peptide is pulled or pushed through the CNT. On the other hand, poly-Ser shows the rel.$\Delta P_a$ values in the range between --100\% and +25\%. Mechanical force not only detaches the peptide from the CNT walls in most situations, but can also lead to an opposite process. Poly-Ala behaves similarly to poly-Ser -- the mechanical force decreases the $P_a$, except the relative values are lower than in the case of poly-Ser. 

The loading rate of the force probe affects the association/dissociation ratio only slightly. The highest loading rate of 20 pN\,ps$^{-1}$ causes more extreme changes in rel.$\Delta P_a$ than the lower loading rates in the three sequences. There appears to be no effect of the type of applied force on rel.$\Delta P_a$.

\begin{figure}[tb]
    \centering
    \includegraphics{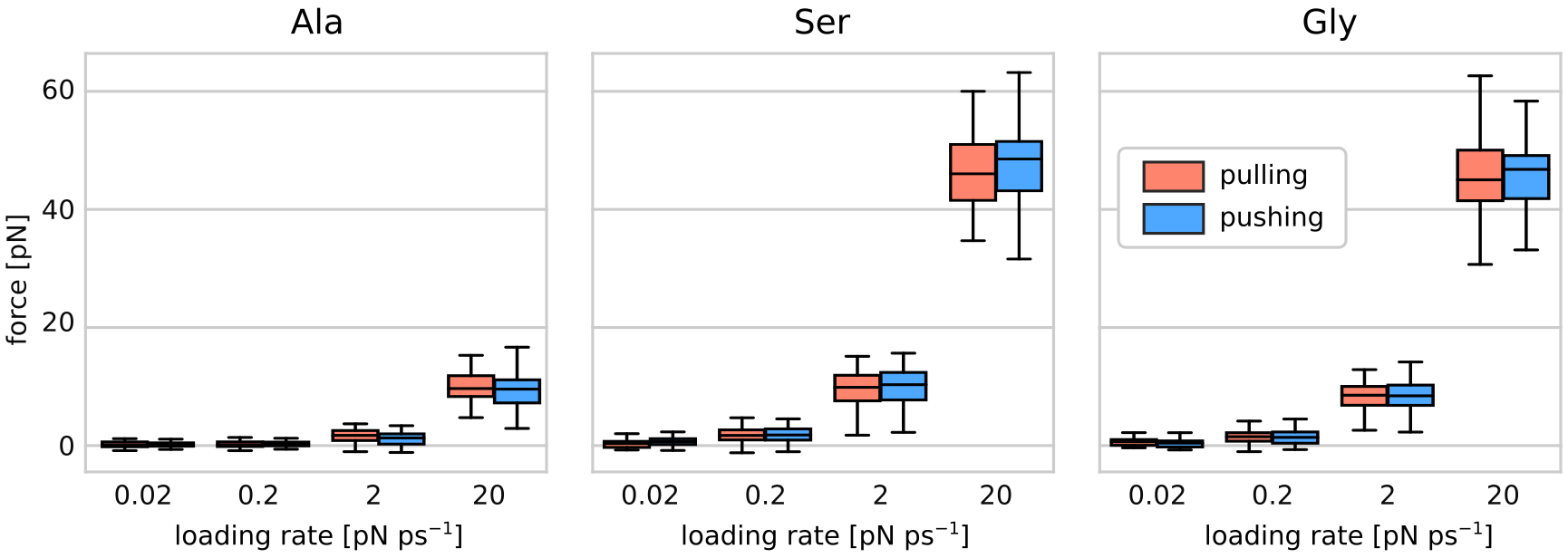}
    \caption{The force acting on the peptide averaged over each fpMD trajectory. The boxes represent interquartile ranges of the per-trajectory averages with the mean values indicated. The whiskers stand for the minimum/maximum values excluding the outliers.}
    \label{fig:forces}
\end{figure}

The fpMD simulations were performed in the constant-velocity regime. Hence, the force acting on the force probe fluctuated extensively. The analysis of the forces is shown in Fig.\,\ref{fig:forces}. We averaged the force over the fpMD trajectories. The positive average force means that the peptide resists movement. We observed that for slower simulations (i.e. low loading rate) the average force acting on the peptide is close to zero. In these cases, the system allows for an effective distribution of the mechanical force. However, for higher loading rates, the average force is non-negligible and averages to about 10\,pN for poly-Ala and 50\,pN for poly-Ser and poly-Gly. These forces are in the range needed to release translational arrest of the ribosome, as shown by optical tweezers experiments \cite{Goldman15}.

\section{Conclusions}

In this work, we investigated how peptides respond to mechanical force in a confined space represented by a CNT. In real life, we are familiar with several macroscopic situations that shape our general intuition. For example, pulling a string from a hoodie is easy, but restringing it requires some skill because the string is flexible and easily gets stuck. Here, using force-probe MD simulations, we found that this intuition may fail in the microscopic world.

Our main goal was to characterize how pulling a peptide through a CNT differs from pushing it in the same direction. A broader motivation involves various cellular processes in which peptides overcome confined regions of biomolecular tunnels or cavities. The nature of the forces that act on the peptides remains unclear but may be important for understanding the mechanism of the processes. Using a model system comprising a carbon nanotube, a decapeptide, and water, we found that the intrinsic properties of the peptides, such as the sequence of amino acid residues, may play a more important role than the way the external force acts on the peptide.

Our simulations revealed that for low loading rates, the internal conformational dynamics \emph{outweight} the external mechanical force. The differences in the conformational dynamics of poly-Ser and poly-Gly, chemically the most distinct polypeptides studied, were greater than the differences caused by different attachment points of the force probe. In general, a stronger effect of external forces was more apparent for higher loading rates, which are, however, less relevant for biological processes that are typically slower than the fastest fpMD simulations in this work.

However, our simulations suggested that the type of applied force, whether it pushes or pulls, plays a role in the relative position of the peptide within CNT. With force action, the peptides tend to dissociate from the CNT wall. Furthermore, pulling the peptide seems to maintain its orientation within the confined space better than pushing it.

The obvious limitation of our approach is the use of homorepeats. In cells, repeats of single amino-acid residues are not rare, especially in eukaryots. The abnormal occurrence of homorepeats is associated with pathologies \cite{Chavali20}. Poly-glutamine stretches, for example, are related to several neurodegenerative diseases \cite{MacDonald93}. Still, our main motivation in this work was to simplify the problem and focus on the force attachment points. Using peptides with nonuniform sequences would bring challenges such as considering acid-base equilibrium under confinement or counterions in the simulation box. These fall outside the scope of our work. 

Furthermore, our results are valid within the scope of chemically uniform CNT. In future studies, a more realistic biological environment may provide an even more realistic picture of how external forces affect the dynamics of peptides or proteins. Our approach provides solid foundations for future force-probe MD simulations of peptide translocations through confined spaces.

\section*{Acknowledgment}

The authors thank J. Cikhart and H. McGrath for critical comments on the manuscript and P. Chalupský and other members of Kolář group for inspiring discussions. This work was supported by the Ministry of Education, Youth and Sports of the Czech Republic through the e-INFRA CZ (ID:90254), and by the Czech Science Foundation (project 23-05557S). This work was also supported by the grant of Specific university research No. A2\_FCHI\_2021\_034, and A1\_FCHI\_2022\_002. 

\section*{Author's contribution}

MHK designed and supervised the research. FCN set up, performed, and analyzed the simulations. Both authors interpreted the results. Both authors prepared the figures. FCN wrote the first version of the manuscript, both authors finalized the manuscript.

\section*{Conflict of interest}

The authors declare no conflict of interest.

\section*{Supplementary information}

The files needed to perform equilibrium and fpMD simulations in GROMACS, the output data, and the analysis script to generate all figures are available at \href{ttps://github.com/mhkoscience/nepomuceno-cnt-peptides}{https://github.com/ mhkoscience/nepomuceno-cnt-peptides}

\providecommand{\latin}[1]{#1}
\makeatletter
\providecommand{\doi}
  {\begingroup\let\do\@makeother\dospecials
  \catcode`\{=1 \catcode`\}=2 \doi@aux}
\providecommand{\doi@aux}[1]{\endgroup\texttt{#1}}
\makeatother
\providecommand*\mcitethebibliography{\thebibliography}
\csname @ifundefined\endcsname{endmcitethebibliography}
  {\let\endmcitethebibliography\endthebibliography}{}

\end{document}